\documentclass[aps,nature,longbibliography,twocolumn,superscriptaddress]{revtex4-1}
\usepackage{amsmath}
\usepackage{amsthm}
\usepackage{mathrsfs}
\usepackage{graphicx}
\usepackage{color}
\usepackage{tabularx}
\usepackage{array}
\usepackage{multirow}
\usepackage{hyperref}

\newcommand{\dee}{\mathrm{d}}

\newcommand{\ket}[1]{|#1\rangle}

\newcolumntype{L}[1]{>{\flushleft\arraybackslash}m{#1}}
\newcolumntype{M}[1]{>{\centering\arraybackslash}m{#1}}

\begin{document}

\title{Spectrally engineering photonic entanglement with a time lens}

\author{J. M. Donohue}
\email[]{jdonohue@uwaterloo.ca}
\affiliation{Institute for Quantum Computing and Department of Physics \&
Astronomy, University of Waterloo, Waterloo, Ontario, Canada N2L 3G1}
\author{M. Mastrovich}
\affiliation{Institute for Quantum Computing and Department of Physics \&
Astronomy, University of Waterloo, Waterloo, Ontario, Canada N2L 3G1}
\affiliation{Department of Physics, Harvey Mudd College, Claremont, California 91711, USA}
\author{K. J. Resch}
\email[]{kresch@uwaterloo.ca}
\affiliation{Institute for Quantum Computing and Department of Physics \&
Astronomy, University of Waterloo, Waterloo, Ontario, Canada N2L 3G1}

\begin{abstract}\noindent A time lens, which can be used to reshape the spectral and temporal properties of light, requires ultrafast manipulation of optical signals and presents a significant challenge for single-photon application.  In this work, we construct a time lens based on dispersion and sum-frequency generation to spectrally engineer single photons from an entangled pair.  The strong frequency anti-correlations between photons produced from spontaneous parametric downconversion are converted to positive correlations after the time lens, consistent with a negative-magnification system.  The temporal imaging of single photons enables new techniques for time-frequency quantum state engineering.\end{abstract}

\maketitle

Sources of single photons with precisely controlled properties are necessary for effective and efficient photonic quantum communication, computation, and metrology.  The spectral, or energy-time, degree of freedom is of particular interest, as it can be used to encode information in a high-dimensional Hilbert space~\cite{nunn2013large} and is naturally robust when transmitting through both long-distance fiber links~\cite{tanzilli2005photonic} and photonic waveguides~\cite{brecht2015photon}.  Entanglement in this degree of freedom is essential for applications such as high-dimensional quantum key distribution~\cite{nunn2013large}, nonlocal dispersion cancellation~\cite{franson1992nonlocal}, and quantum-enhanced clock synchronization~\cite{giovannetti2001quantum}.  The nonlinear process of spontaneous parametric downconversion (SPDC), for example, provides a reliable source of energy-time entangled photons.  Because of energy conservation, most SPDC sources tend to produce photons with frequency anti-correlations.  However, photon pairs with positively correlated spectra may be useful for dispersion cancellation in long-distance channels~\cite{lutz2014demonstration} and quantum-enhanced clock synchronization~\cite{giovannetti2001quantum}.  Using SPDC sources with extended phasematching conditions, joint spectra with positive spectral correlations have been produced before~\cite{kuzucu2008joint,eckstein2011highly,ansari2014probing,lutz2014demonstration}, but control over the correlations after state generation has not yet been demonstrated.


\begin{figure}[b!]
  \begin{center}
           \includegraphics[width=1\columnwidth]{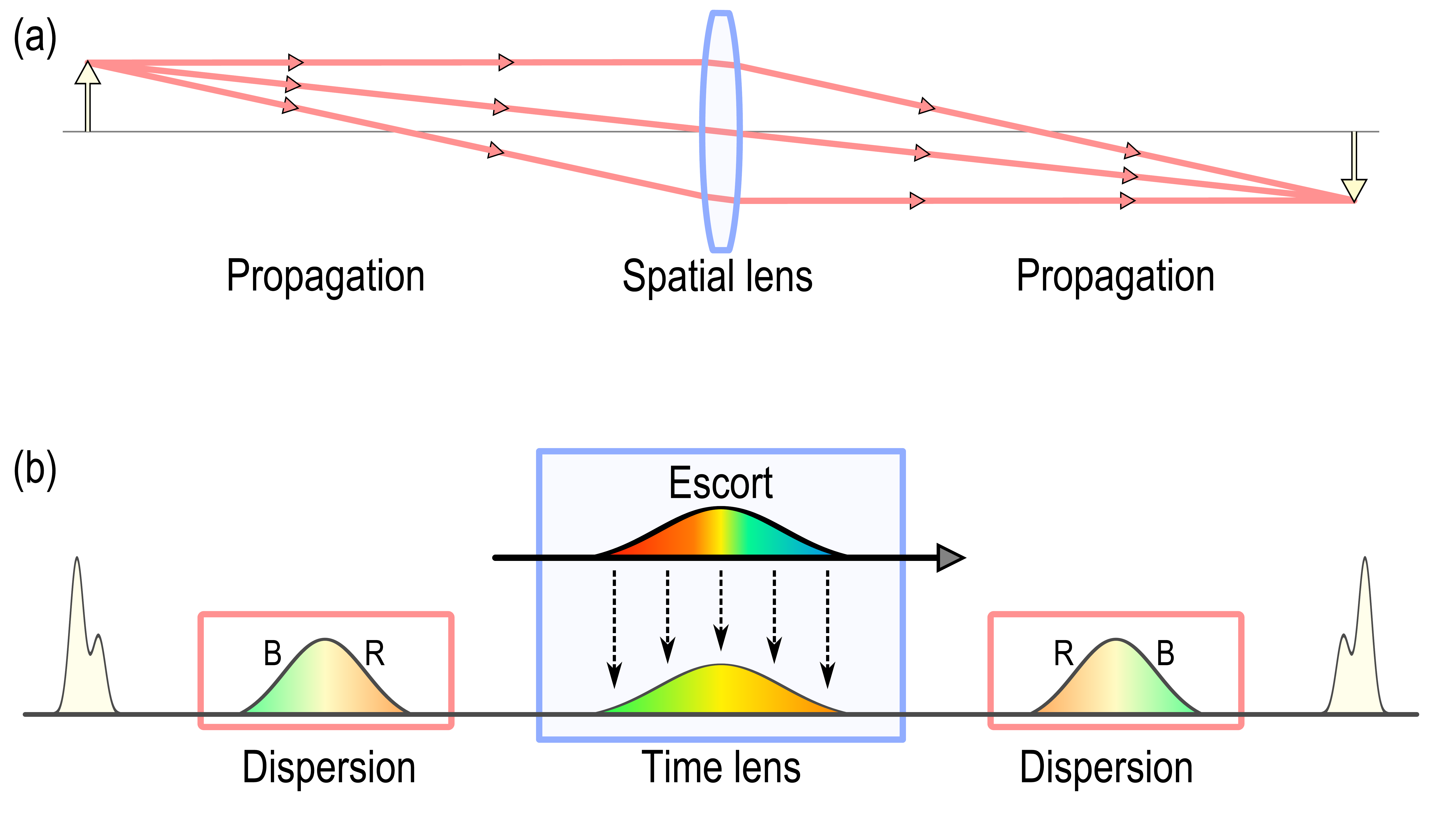}
  \end{center}
  \vspace{-0.8cm}
 \caption{\textbf{Temporal shaping with an upconversion time lens.} (a) In a spatial imaging system, free-space propagation spreads the spatial extent of the beam such that each portion of the beam has a distinct transverse momentum, visualized with arrows.  The lens shifts the momenta in a spatially dependent fashion, which effectively reverses the momenta for off-centre components, and the beam refocuses with further spatial propagation.  (b)  The temporal imaging system operates through an analogous principle, where chromatic dispersion spreads the temporal profile of the beam such that each temporal slice of the beam has a distinct central frequency, ranging from a red-shifted leading edge, R, to a blue-shifted tail, B.  The time lens introduces a time-dependent frequency shift, which can reverse the frequency shifts and allow the wavepacket to refocus itself after more chromatic dispersion is applied. The temporal structure of the pulse will be reversed, akin to an imaging system with negative magnification.  In our realization, we use sum-frequency generation with a dispersed escort pulse to implement an upconversion time lens. At each time in the interaction, the signal interacts with a different frequency of the escort, effectively enforcing a time-dependent relative frequency shift as well as a change in carrier frequency.}\label{fig:concept}
\end{figure}

With linear optics, it is possible to shape the spectrum of single photons through frequency-dependent attenuation and phase shifts, as can be accomplished with pulse shapers and spatial-light modulators~\cite{pe2005temporal}.  More complex transformations demand techniques such as those based on fast electro-optic devices or nonlinear optical processes.  These are required for many quantum applications, such as interfacing with quantum memories~\cite{albrecht2014waveguide}, ultrafast photon switching~\cite{hall2011ultrafast}, manipulating time-bin qubits~\cite{humphreys2013linear,donohue13,nowierski2015experimental}, and temporal mode selection~\cite{eckstein2011quantum,brecht2015photon}.   Spectral control over a photon after it has been created is therefore highly desirable for ultrafast manipulation and state engineering, especially at wavelengths where materials with suitable phasematching do not exist.  Entanglement cannot be increased or decreased through a local unitary process.  As a result, pulse shaping cannot be used to convert energy-time entangled states to uncorrelated states~\cite{grice2001eliminating}.  It can, however, be used to change the specific energy relations between the two photons or convert the correlation between the photon frequencies to a correlation between the frequency of one and the time of arrival of the other.

In this work, we demonstrate ultrafast control of quantum-optical waveforms with sub-picosecond features.  We construct a temporal imaging system based on ultrafast nonlinear effects to manipulate the spectral profile of single photons.  We apply this technique to half of an energy-time-entangled pair produced with SPDC, and observe that the frequency anti-correlations are converted to positive correlations after the time lens through joint spectral  intensity measurements, in addition to an adjustable central frequency shift.  It is straightforward to adjust the temporal spectral magnification by changing the chirp parameters, and our scheme is free of intense broadband noise such as Raman scattering.  A similar, as-yet-unrealized application of time lenses was proposed using electro-optic modulators~\cite{tsang2006propagation}.

Temporal imaging can be understood in direct analogy with its spatial counterpart~\cite{kolner1994space,walmsley2009characterization}.  It is instructive to compare the corresponding elements in each, as shown in Fig.~\ref{fig:concept}.  In spatial imaging, free-space propagation causes the momentum components to diverge in space, resulting in a spatial spread of the beam.  The action of the lens is to shift the transverse momenta in a spatially dependent way, such that after further propagation the beam may refocus.  Analogously, in temporal imaging, propagation through a dispersive material (such as optical fiber) causes the constituent frequencies of a pulse to diverge in time, resulting in a temporal spread of the pulse.  Constructing the equivalent of a lens for temporal imaging requires a time-dependent frequency shift in the same way that a spatial lens requires a spatially dependent transverse momentum shift.  Self-phase modulation can approximate the required effect~\cite{walmsley2009characterization}, but is ineffective for single-photon signals. Four-wave mixing has been shown to be effective for classical signals~\cite{jopson1993compensation,watanabe1993compensation,foster2008silicon,salem2008optical,foster2009ultrafast,gu2013demonstration}, but suppression of broadband noise sources presents a challenge for quantum signals. Recent work has shown great promise using cross-phase modulation in photonic crystal fiber~\cite{matsuda2016deterministic}, Raman memories~\cite{fisher2016frequency}, and electro-optic modulators~\cite{lukens2015electro,karpinski2016bandwidth} to shape broadband single-photon waveforms.

The upconversion time lens~\cite{bennett1994temporal}, seen in Fig.~\ref{fig:concept}b, is based on sum-frequency generation (SFG), a type of three-wave mixing in which two pulses may combine to produce a pulse at the sum of their frequencies.  In the case of interest, one pulse is considered to contain a single photon and the other to be a strong classical pulse.  This strong classical pulse, referred to as the \emph{escort}, upconverts the photon to a new frequency, leaving its own imprint on the spectral shape of the upconverted photon.  SFG has been shown to be a powerful and potentially efficient tool for ultrafast waveform manipulation which remains effective at the single-photon level~\cite{Huang1992FreqConv,kwiat_upconversion,tanzilli2005photonic,agha2014spectral,donohue2015theory,clemmen2016ramsey}.  The combination of SFG and pulse shaping enables the manipulation of ultrafast single-photon waveforms for techniques such as bandwidth compression~\cite{lavoie13comp}, quantum pulse gates~\cite{eckstein2011quantum}, and time-to-frequency conversion~\cite{foster2008silicon,donohue13}.

\begin{figure}[t]
  \begin{center}
         \includegraphics[width=1\columnwidth]{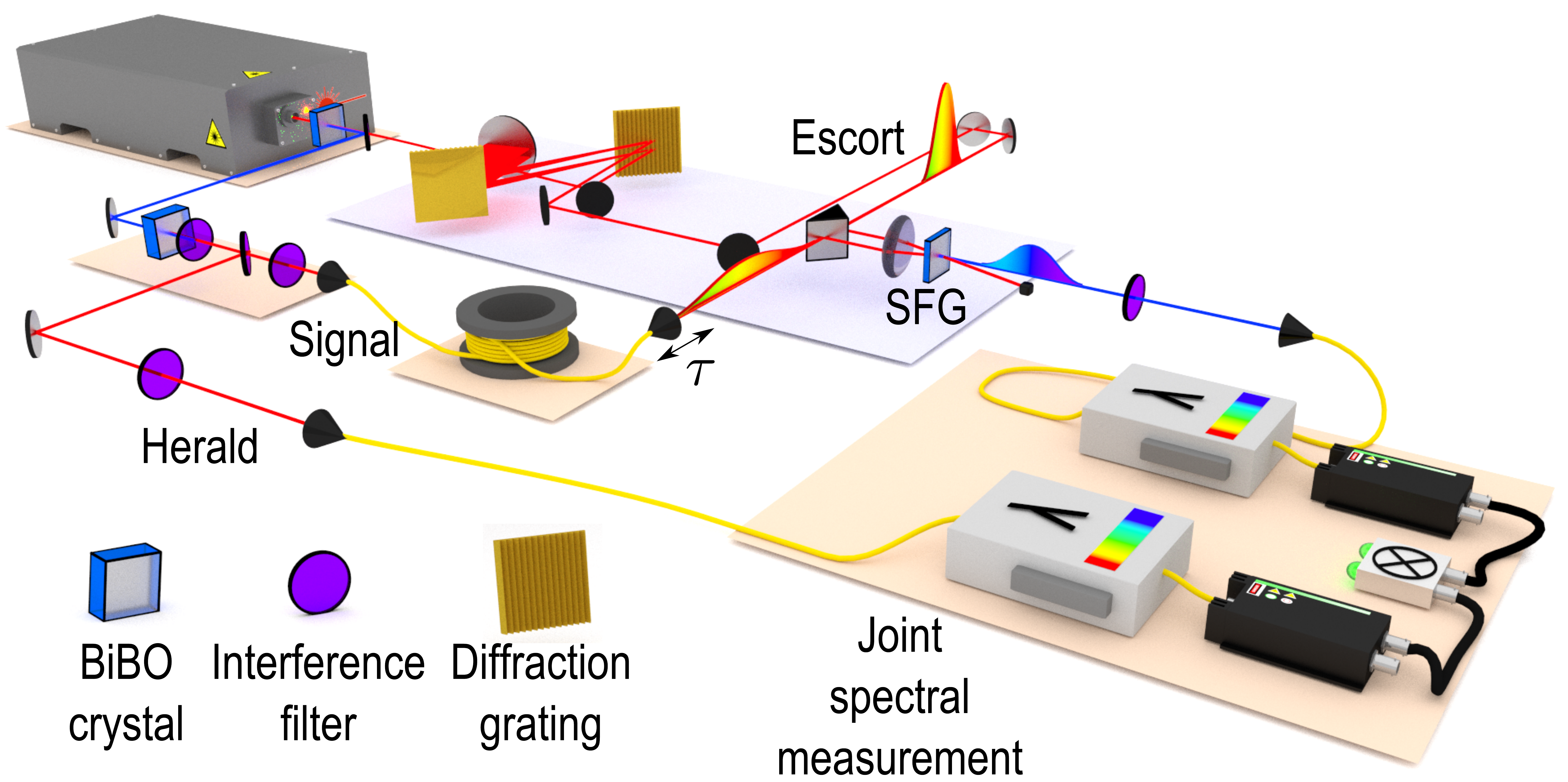}
  \end{center}
    \vspace{-0.3cm}
 \caption{\textbf{Experimental setup.} Frequency-entangled photons are created through the spontaneous parametric downconversion of ultrafast pulses from a frequency-doubled Ti:Sapph laser.  The signal photons are chirped through 34~m of single-mode fiber, while the remaining Ti:Sapph light comprises the escort pulse and anti-chirped in a grating-based pulse compressor.  The pulses are recombined with relative delay $\tau$ for non-collinear sum-frequency generation (SFG). The upconverted signal is then isolated with bandpass filters and spectrally resolved in coincidence with the herald.}\label{fig:setup}
\end{figure}

We represent the spectral field of a dispersed optical pulse as $F(\omega)e^{i\phi(\omega)}$, where the spectral phase has a quadratic frequency dependence, ${{\phi(\omega)}={A(\omega-\omega_0)^2}}$, with chirp parameter $A$.  We characterize the dispersion applied to the input signal, escort pulse, and output waveform by the chirp parameters $A_i$, $A_e$, and $A_o$, respectively, which in the case of normal dispersion are proportional to the length of material passed through.  We assume that the pulses are all chirped to many times their initial widths, known as the large-chirp limit. In this limit, the imaging equation for the time lens system shown in Fig.~\ref{fig:concept}b can be found in Refs.~\cite{bennett1994temporal,donohue2015theory} and simplifies to the relation~\cite{walmsley2009characterization,donohue2015theory}, \begin{equation}\frac{1}{A_{i}}+\frac{1}{A_{o}}=-\frac{1}{A_{e}}.\end{equation}   This equation has the same form as the thin lens equation, with dispersion playing the role of propagation distance and the escort chirp the role of the focal length.

In analogy to spatial imaging, the output waveform will have the same features as the input but scaled by a magnification factor~\cite{bennett1994temporal,walmsley2009characterization,donohue2015theory} \begin{equation}M_{\mathrm{spectral}}=\frac{1}{M_{\mathrm{temporal}}}=\left(-\frac{A_{i}}{A_{o}}\right)=1+\frac{A_{i}}{A_{e}}.\label{eq:magnification}\end{equation}  The inverse relationship of the spectral and temporal magnification is a consequence of the scaling property of the Fourier transform. Because of this relationship, spectral measurements can be used to observe the time lens effect.

If $A_{i}=-2A_{e}$, the effective temporal and spectral magnification is $-1$, and both the temporal and spectral shapes will be reversed.  If the input signal is a single photon which is spectrally entangled with a partner, reversing the spectrum of the photon will result in an overall reversal of the two-photon joint spectrum.  The reversal of the spectral profile occurs as the photonic signal is chirped twice as strongly as the escort, such that every portion of the signal spectrum detuned from the central frequency by $\delta\omega$ meets a segment of the escort pulse detuned by $-2\delta\omega$ from its central frequency.  In the large-chirp limit, only an output chirp $A_o$ is required to recompress the joint spectral state temporally, which has no effect on the spectral profile of the output but determines the output joint temporal distribution.   A derivation may be found in Supplementary Section I.

\begin{figure}[t!]
  \begin{center}
           \includegraphics[width=1\columnwidth]{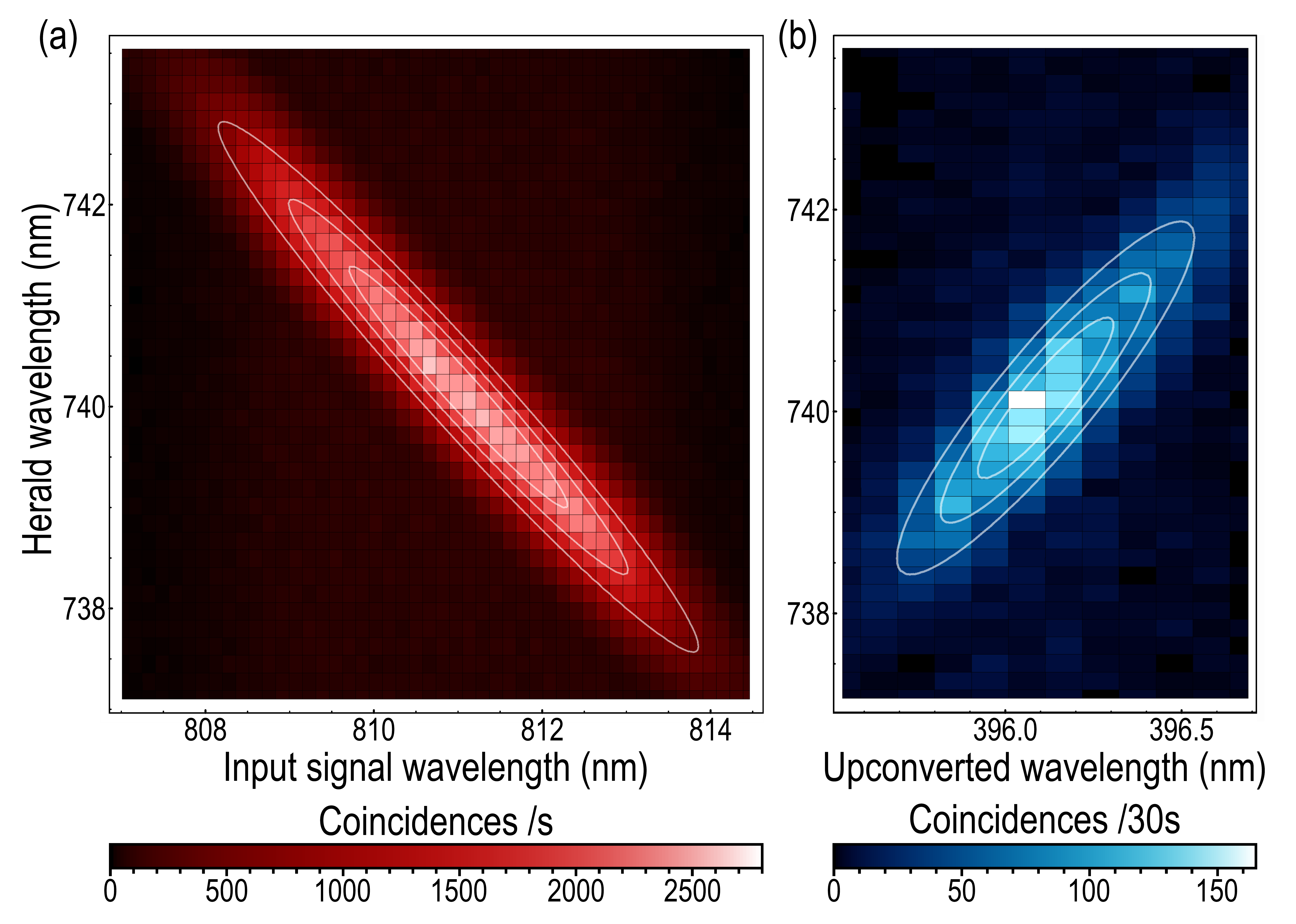}
  \end{center}
    \vspace{-0.3cm}
 \caption{\textbf{Joint spectra.} (a) The joint spectrum measured between the signal photon and the herald immediately after downconversion has strong frequency anti-correlations, exhibiting a statistical correlation of $-0.9702\pm0.0002$ ($-0.9776\pm0.0009$ when corrected for finite spectrometer resolution).  (b) After sum-frequency generation, the joint spectrum between the upconverted signal photon and the herald exhibits strong positive frequency correlations, with a statistical correlation of $+0.863\pm0.004$ ($+0.909\pm0.005$ resolution-corrected).  The white lines on each plot correspond to 25\% contours of the resolution-corrected Gaussian fits.  Background subtraction has not been employed in either image.}\label{fig:jointspec}
\end{figure}

\begin{table}[b]\centering
\begin{tabular}{|M{0.31\columnwidth}|M{0.33\columnwidth}|M{0.33\columnwidth}|}
\hline
 Property & Input & Output \\
 \hline\hline
 Signal bandwidth & $(1.840\pm0.003)$~THz & $(1.14\pm0.02)$~THz\\ \hline
 Herald bandwidth & $(2.034\pm0.003)$~THz & $(1.35\pm0.02)$~THz\\ \hline
 Correlation $\rho$  & $-0.9776\pm0.0009$ & $0.909\pm0.005$\\ \hline
 $g^{(2)}_{s,h}$ & $4.190\pm0.002$ & $3.34\pm0.03$\\ \hline
\end{tabular}
\caption{\textbf{Joint spectral state parameters.}  Selected properties of the Gaussian fits to the joint spectra seen in Fig.~\ref{fig:jointspec} are given above.  The values are all corrected for the finite resolution of the spectrometers, and the widths are full-width at half-maximum.   The bandwidth of the herald photon is reduced when measured in coincidence with the upconverted signal due to the finite temporal width of the chirped escort and the frequency correlations in the initial joint spectrum.  The correlation parameters show that the spectral correlation is present after the time lens, but has changed from anti-correlation to positive correlation.  The second-order cross-correlation $g^{(2)}_{s,h}$ is significantly greater than 2, indicating nonclassical statistics.\label{tab:jointspec} }
\end{table}

Our experimental setup is shown in Fig.~\ref{fig:setup}, and is detailed in Supplementary Section II.  We create signal and herald photons through SPDC in 3 mm of bismuth borate (BiBO) stimulated by a pulsed laser.  The downconverted photons are spectrally filtered and coupled into single-mode (SM) fiber, from which they can each be sent to independent grating-based scanning spectrometers for spectral analysis, as seen in the initial joint spectrum of Fig.~\ref{fig:jointspec}a.  The signal photons can be instead sent through 34~m of SM fiber, where they are positively chirped by normal dispersion. The escort laser pulse passes through a grating-based compressor, set to apply half the dispersion as the fiber with the opposite sign~\cite{lavoie13comp}.  The escort and signal photon are combined for SFG in 1 mm of BiBO, and the upconverted photon is measured with another scanning spectrometer in coincidence with the herald.  The combined efficiency of the chirp, upconversion, and fiber coupling is approximately 0.2\%.  Detection events from the herald spectrometer were measured in coincidence with the upconverted photon spectrum, and the measured joint spectrum is shown in Fig.~\ref{fig:jointspec}b.

The output joint spectrum exhibits clear positive frequency correlations, in contrast with the clear anti-correlations seen in the input joint spectrum. The measured joint spectra were then fit to a two-dimensional Gaussian.  All spectrometers used had a spectral resolution of approximately 0.1~nm.  While this resolution allowed us to resolve the essential spectral features, this finite resolution is on the same order of magnitude as our spectral bandwidths, which broadened their measured features.  To account for the limited resolution, we deconvolved the fit spectra with a Gaussian spectrometer response function.  The fit parameters of the joint spectra produced from our SPDC source and after SFG are shown in Table~\ref{tab:jointspec}. The statistical correlation of the signal and herald wavelengths $\rho$, defined as the covariance of two parameters divided by the standard deviation of each, is statistically significant both before and after the time lens. The negative-to-positive change of $\rho$ indicates the reversal of the correlations.

Assuming that the two-photon state is pure and coherent, we have succeeded in manipulating energy-time entanglement.  Previous experiments have established that photons created through SPDC are energy-time entangled rather than classically correlated~\cite{kwiat93fringes} and that SFG maintains coherence~\cite{tanzilli2005photonic,donohue13,clemmen2016ramsey}.  Modeling our experiment under these assumptions, our data show that the photon pairs have been converted from energy-time-entangled states with strong frequency anti-correlations to ones with strong frequency correlations.  The temporal distribution is not uniquely specified by the joint spectrum as the spectral phase remains unknown.  However, if the state is indeed pure, the spectral phase could be compensated to make a transform-limited distribution with tight temporal anti-correlations.   To directly measure the time-domain distribution requires time resolution on the order of femtoseconds, which is much shorter than the resolution of our photon counters.  The timescale of our experiment could be measurable with additional nonlinear processes~\cite{kuzucu2008joint}, and energy-time entanglement could be directly confirmed through a nonlocal interference experiment with additional temporal selection~\cite{franson89}.

\begin{figure}[t!]
  \begin{center}
           \includegraphics[width=1\columnwidth]{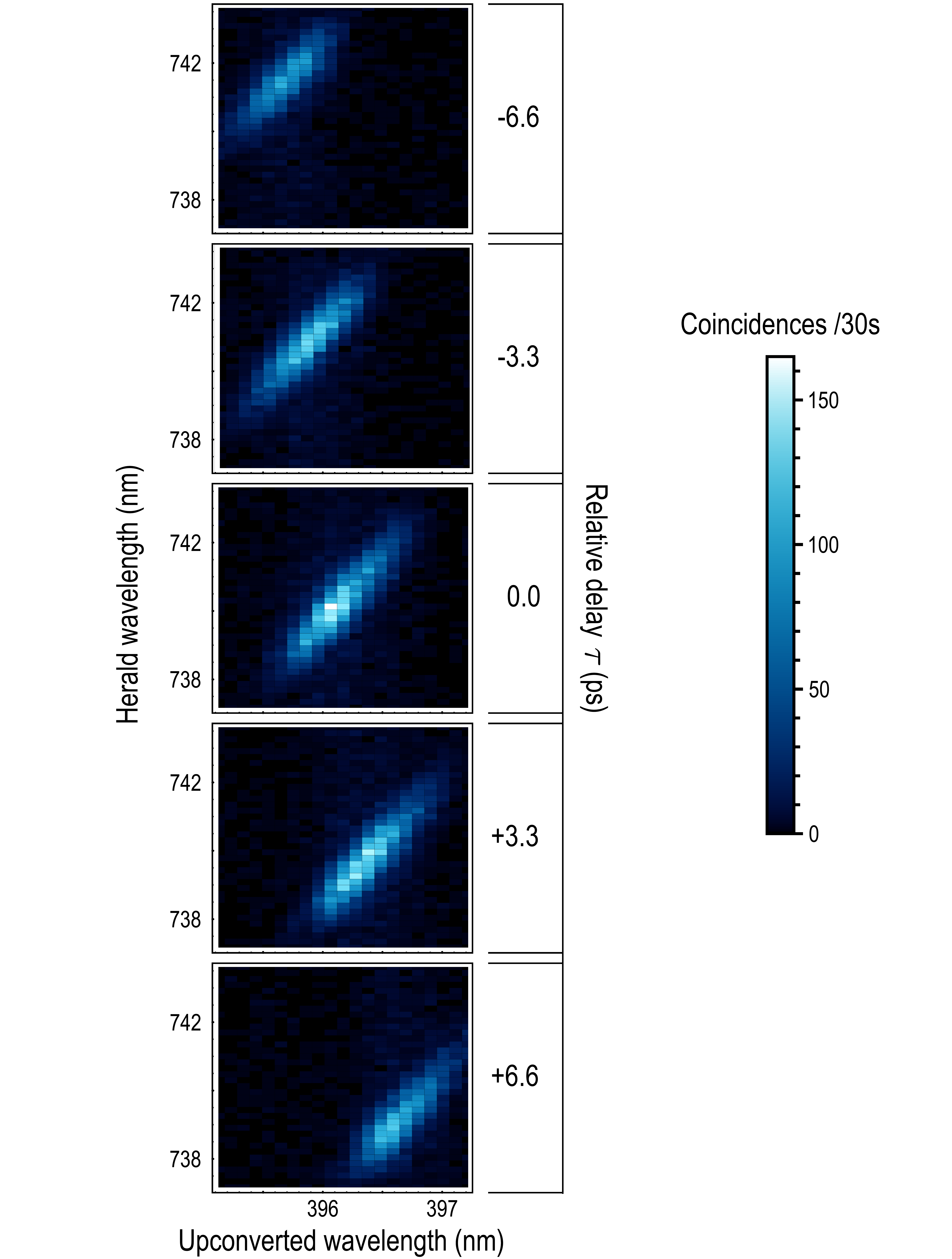}
  \end{center}
  \vspace{-0.25cm}
 \caption{\textbf{Tunability of the joint spectra.} The center frequency of the joint spectrum is tunable by introducing a relative delay between the input signal and the escort pulse, as seen in the five measured joint spectra.  The central wavelength of the upconverted signal changes by 0.071~nm per picosecond of delay.  The herald central frequency is also seen to shift due to the escort acting partially as a temporal filter on the signal.}\label{fig:jointspectune}
\end{figure}

We also calculate the second-order cross-correlation function between the signal and herald photon $g_{s,h}^{(2)}=\frac{P_{s\& h}}{P_{s}P_h}$ when the two photons are coincident in time, where $P_i$ is the probability of measuring a photon in mode $i$, by comparing the coincidences with the single-detection events before the spectrometers~\cite{albrecht2014waveguide,fisher2016frequency}. A value larger than two indicates nonclassicality if we assume that the individual second-order statistics of the signal and herald are at most thermal (as expected for a single-photon state)~\cite{fisher2016frequency}. The values of the second-order cross-correlation function of the herald with the initial and upconverted signal photon are given in Table~\ref{tab:jointspec}, and are significantly larger than two in both cases.  The lower $g_{s,h}^{(2)}$ value of the upconverted light can be attributed to the spectrally distinguishable uncorrelated second-harmonic background which reaches the detectors. An extensive list of measured parameters, including other correlation quantifiers, may be found in Supplementary Section II.

While the upconversion efficiency of the time lens in our experiment is low, the limitations are practical rather than fundamental~\cite{donohue2015theory}.  As the escort used in our experiment is approximately the same spectral width as the photons and is only chirped half as much, its chirped temporal duration does not fully envelop the signal photon. As such, it acts partially like a temporal filter, evidenced by the reduction in the herald bandwidth and statistical correlation of the final joint spectrum.  An escort pulse with a significantly broader spectrum than the photon would increase the efficiency of the process.  Further efficiency concerns could be addressed with higher power escort pulses and materials with stronger nonlinearities, although phasematching restrictions must be carefully considered.

By analogy to a light beam incident off centre to a lens, introducing a relative delay between the escort pulse and input signal results in a shift in the central wavelength of the joint spectrum, as shown in Fig.~\ref{fig:jointspectune}.  This occurs as the relationship between the instantaneous frequencies of the chirped signal and anti-chirped escort are shifted, resulting in an overall central frequency shift.  By tuning the relative delay over a range of $\sim$13~ps, the central frequency can be tuned over a range of $\sim$2~THz. However, it is seen that the central wavelength of the herald also changes as the delay is changed, as the upconversion time lens does not uniformly support the entire bandwidth of the input photon.  Once again, ensuring that the escort has a broader bandwidth would solve this problem, and allow for high-efficiency tunability over a wide spectral range~\cite{donohue2015theory}.  Restrictive phasematching of the upconversion medium will also result in a narrowing of the upconverted signal spectra.  These effects are detailed in Supplementary Section III.

We have demonstrated control of a twin-photon joint spectral intensity through the use of an upconversion time lens on the ultrafast timescale.  The technique presented maintains two-photon correlations and introduces minimal background noise at the target wavelengths.  The time lens demonstrated here is an essential component of a general quantum temporal imaging system, capable of essential tasks such as bandwidth compression, time-to-frequency conversion, and all-optical transformations on time-bin-encoded qubits.  Control of the correlation of joint spectra as demonstrated here can be used to create spectrally correlated two-photon states at wavelengths where efficient nonlinear materials with extended phasematching conditions do not exist~\cite{grice2001eliminating,kuzucu2008joint,eckstein2011highly,lutz2014demonstration}. Such a technique may be directly useful for shaping the spectra of entangled pairs for long-distance communications and quantum-enhanced metrology~\cite{giovannetti2001quantum} and, more generally, to mold the time-frequency distributions of single photons for experiments both fundamental and practical.

The authors thank J.~Lavoie, M.~Karpi\'{n}ski, M.~D.~Mazurek, and K.~A.~G.~Fisher for fruitful discussions, and C. Mastromattei for valuable assistance in the laboratory.  We are grateful for financial support from the Natural Sciences and Engineering Research Council, the Canada Foundation for Innovation,  Industry Canada, Canada Research Chairs, and the Ontario Ministry of Research and Innovation.



\renewcommand{\thesubsection}{\thesection.\arabic{subsection}}
\makeatletter 
\def\tagform@#1{\maketag@@@{(S\ignorespaces#1\unskip\@@italiccorr)}}
\makeatother
\makeatletter
\makeatletter \renewcommand{\fnum@figure}
{\figurename~S-\thefigure}
\makeatother
\makeatletter
\makeatletter \renewcommand{\fnum@table}
{\tablename~S-\thetable}
\makeatother
\renewcommand{\figurename}{Figure}
\setcounter{equation}{0}
\setcounter{figure}{0}
\setcounter{table}{0}

\onecolumngrid

\section*{Supplementary Material}

The supplementary material is organized as follows. Firstly, in Section~\ref{sec:theory}, we flesh out the analogy between a spatial lens and a time lens. We then use a first-order perturbative method to model the upconversion time lens, and show how to extract the magnification factor. In Section~\ref{sec:TLExpDetails}, we provide the details of the experimental setup to supplement the main text as well as additional fit parameters of the measured joint spectra.    In Section~\ref{sec:tunability}, we explore the effect of a relative time delay between the escort and the input signal, and show how it can be used to both adjust the center wavelength of the final signal and evaluate the effectiveness of the time lens itself.

\section{Detailed theoretical description}\label{sec:theory}

\begin{figure}[h]
  \begin{center}
           \includegraphics[width=1\columnwidth]{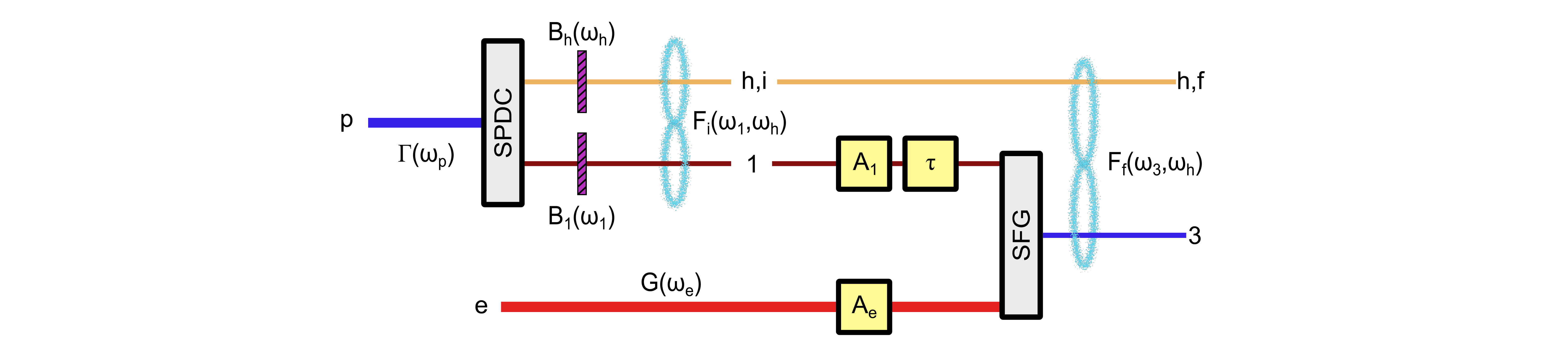}
  \end{center}
 \caption{\textbf{Abstract representation of setup.}  The process under study begins with spontaneous parametric downconversion (SPDC) of a pump $p$ with spectrum $\Gamma(\omega_p)$.  The photon pair is filtered with bandpass filters $B_j(\omega_j)$.  The photon pair, with a signal in mode $1$ and a herald in mode $h$, share a joint spectral amplitude of $F_i(\omega_1,\omega_h)$.  The signal is chirped, represented by the parameter $A_1$. An escort pulse in mode $e$ with spectrum $G(\omega_e)$ is also chirped, represented by the parameter $A_e$, as well as delayed in time relative to the input signal by an amount $\tau$.  The signal photon and escort are mixed for sum-frequency generation (SFG), and the output joint spectral amplitude between the output signal in mode $3$ and the herald is $F_f(\omega_3,\omega_h)$.}\label{fig:subscripts}
\end{figure}

In this section, we explain the upconversion time lens. We first provide intuition for its operation by relating the quadratic phases of a spatial lens to those of dispersion and upconversion. We then tackle a Gaussian system with a first-order perturbative method.  We consider solely the low-efficiency regime, where we can safely ignore effects of over-conversion and time-ordering~\cite{donohue2015theory,brecht2015photon}. For ease of reference to subscripts and notation, see Fig.~S-\ref{fig:subscripts}.

\subsection{Contrasting a spatial lens and a time lens}

As a Gaussian beam propagates in free space, it's spatial distribution spreads out but its momentum distribution (given by the Fourier transform) remains constant.  Solving the monochromatic paraxial Helmholtz equation, while ignoring $y$-dependence for simplicity, leads to the relation~\cite{kolner1994space} \begin{equation}F(k_x,z)=F(k_x,0)e^{i\frac{z}{2k_0}k_x^2},\label{eq:spatprop}\end{equation} where $k_0=n\omega_0/c$ is the $z$-momentum in the paraxial approximation. While the spatial intensity distribution $f(x,z)$ expands, this quadratic phase does not change the distribution of transverse momenta, $|F(k_x,z)|$.  As the beam propagates, the momentum components spread in space such that, in the large-$z$ limit, they each occupy a unique point; this spreading can therefore be thought of as a momentum-dependent spatial shift.

In contrast, a perfect parabolic lens in the thin-lens approximation imparts a phase that varies quadratically in space, transforming the spatial field to~\cite{kolner1994space} \begin{equation}f(x,z)\mapsto f(x,z)e^{i\frac{k_0}{2f}x^2},\label{eq:spatlens}\end{equation} where $f$ is the focal length. At the lens, the spatial profile does not change, but it does cause a spatially dependent momentum shift which allows for refocusing and manipulation of beams.  We know~\cite{kolner1994space,walmsley2009characterization} that phases of this form lead to the thin lens equation, \begin{equation}\frac{1}{z_1}+\frac{1}{z_2}=\frac{1}{f},\end{equation} with a magnification \begin{equation}M_{\mathrm{spatial}}=-\frac{z_2}{z_1}.\end{equation}  Since magnification of a Fourier transform obeys the scaling relationship $\mathcal{F}\left[f(ax)\right]\propto F(k_x/a)$, the magnification of the momentum distribution is simply $1/M_{\mathrm{spatial}}$.

We next compare the treatment of a broadband pulse travelling as a plane wave through a medium.  If we once again denote $z$ as the propagation direction, the pulse centred about $\omega_0$ will accumulate phase as \begin{equation}F(\omega,z)=F(\omega,0)e^{ik(\omega)z}=F(\omega,0)e^{ik_0z+ik'(\omega-\omega_0)z+i\frac{k''z}{2}(\omega-\omega_0)^2},\end{equation} where we have expanded the wavevector $k(\omega)$ to second order about the central frequency $\omega_0$. The first-order term corresponds to a time delay, while the second-order term corresponds to group velocity dispersion.  In an analogous fashion as Eq.~\eqref{eq:spatprop}, this quadratic phase will leave the intensity distribution in frequency $|F(\omega)|$ unaltered, but will cause a frequency-dependent temporal shift. This leads to the creation of a chirped pulse, with the strength of chirp represented by the chirp parameter $A$, which is $\frac{k''L}{2}$ for a medium of length $L$.

The missing piece of the puzzle is the temporal equivalent of a lens, an optic which implements the phase \begin{equation}f(t,z)\mapsto f(t,z)e^{iBt^2},\label{eq:timelensdesire}\end{equation} where we can define a temporal focal length as $f_t=\frac{\omega_0}{2B}$. With this in place, we could write analogous thin-lens-type imaging equations as \begin{equation}\frac{1}{A_{in}}+\frac{1}{A_{out}}=4B.\end{equation} Some ways to impart the quadratic temporal phase include electro-optic modulators, self-phase modulation, sum-frequency generation, four-wave mixing, and cross-phase modulation~\cite{walmsley2009characterization}.  Here, we briefly provide an intuitive picture for sum-frequency generation (SFG) with a chirped escort pulse~\cite{bennett1994temporal}.

In SFG, two pulses interact in a medium with a second-order nonlinearity.  One pulse is treated as the weak input represented with the spectro-temporal amplitude $f_i(t)$, and the other as a strong undepleted escort pulse with amplitude $g(t)$.  If the material is lossless and has a phasematching function much broader than the pulses in question, the upconverted temporal waveform in the low-efficiency regime will be given by~\cite{donohue2015theory} \begin{equation}f_f(t)\overset{\mathrm{low-eff.}}{\propto}f_i(t)g(t).\end{equation}

If the escort pulse $g(t)$ is a chirped Gaussian pulse with an envelope described by \begin{equation}G(\omega_e)=\frac{1}{(2\pi)^{1/4}\sqrt{\sigma_e}}\exp\left[-\frac{(\omega_e-\omega_{0e})^2}{4\sigma_e^2}\right]\exp\left[iA_e(\omega_e-\omega_{0e})^2\right],\label{eq:escortpulse}\end{equation} then its temporal description can be written as \begin{equation}g(t)=\left(\frac{2}{\pi}\right)^{\frac{1}{4}}\sqrt{\frac{\sigma_e}{1-4iA_e\sigma_e^2}}\exp\left[-\frac{\sigma_e^2t^2}{(1+16A_e^2\sigma_e^4)}-i\frac{4A_e\sigma_e^4t^2}{1+16A_e^2\sigma_e^4}\right]=g'(t)\exp\left[-i\frac{4A_e\sigma_e^4t^2}{1+16A_e^2\sigma_e^4}\right].\end{equation} If we look at the large-chirp limit (LCL), where the pulse is chirped to many times its initial width ($A_e^2\sigma_e^4\gg1$), we can simplify the representation of the field to \begin{equation}g(t)\approx g'(t)\exp\left[-i\frac{t^2}{4A_e}\right].\end{equation}  If we can also assume that the chirped escort pulse is much longer in time than the input, then $|f_i(t)g(t)|\propto |f_i(t)|$ and the upconverted signal will only gain a phase (and central frequency shift) from the SFG process as \begin{equation}f_f(t)\propto f_i(t)\exp\left[-i\frac{t^2}{4A_e}\right].\end{equation}  Thus, in these limits, SFG implements a temporal phase of $B=-\frac{1}{4A_e}$.  Using this upconversion time lens, the imaging equations take the natural form \begin{equation}\frac{1}{A_i}+\frac{1}{A_o}=-\frac{1}{A_e},\end{equation} with a temporal magnification of \begin{equation}M_{\mathrm{temporal}}=-\frac{A_o}{A_i}=\left(1+\frac{A_i}{A_e}\right)^{-1}.\end{equation}  To obtain these, we had to assume freedom from phasematching and a sufficiently large chirp limit.  Note that increasing the escort chirp decreases the effective focusing power of the time lens, but increases its temporal aperture. In the next section, we remove some of these simplifications and investigate our specific case more deeply.

A general imaging system requires more than a lens with free-space on either side.  For magnifying beams, a full telescope is needed, which consists of at least two free-space propagation steps and two lens. Therefore, two time lens are needed for general reshaping of two-photon spectral states~\cite{foster2009ultrafast}.  However, this can be relaxed in some limiting cases. We will see in the next section that the correlation reversal as shown in the main text can be done with one lens so long as we work in the large-chirp limit. The spatial equivalent of this scenario is working the limit such that the distance between the image plane and the lens is much great than the transverse size of the beam. As the focal distance is increased, recollimation at the image plane after the lens has increasingly little effect. This large-chirp limit must be met in order for spectro-temporal reshaping to be effective with a single time lens, even if that time lens is idealized.

\subsection{First-order quantum sum-frequency generation with chirped pulses}

We begin by describing the joint spectral state generated by SPDC in the low-gain regime post-selected for coincidences (i.e. a single photon pair) as \begin{align}\ket{\psi_i(t)}&=\frac{1}{2\pi}\iint\dee\omega_1\dee\omega_h F_i(\omega_1,\omega_h)\ket{\omega_1}_1\ket{\omega_h}_h,\end{align} where ${\ket{\omega_j}=\hat{a}^\dag_{\omega_j}e^{i\omega_jt}\ket{0}}$ represents a single photon of frequency $\omega_j$ which is in a single spatial/polarization mode.  The initial joint spectral representation $F_i(\omega_1,\omega_h)$ is a function of the pump spectrum $\Gamma(\omega_p)$, the phasematching function ${\Phi_{SPDC}(\omega_1,\omega_h,\omega_p)}$, and any bandpass filters applied $B_j(\omega_j)$~\cite{grice2001eliminating}.  If we assume that each of these functions is individually well-approximated by a Gaussian, we can approximate the joint spectral wavefunction as approximated by a two-variable Gaussian, \begin{align}F_i(\omega_1,\omega_h)&=\Gamma(\omega_1+\omega_h)\Phi_{SPDC}(\omega_1,\omega_h)B_1(\omega_1)B_h(\omega_h)\\&=\frac{1}{\sqrt{2\pi\sigma_1\sigma_{h,i}}(1-\rho_i^2)^{1/4}} \exp\left[{\frac{1}{1-\rho_i^2}\left(-\frac{(\omega_1-\omega_{01})^2}{4\sigma_1^2}-\frac{(\omega_h-\omega_{0h,i})^2}{4\sigma_{h,i}^2}-\frac{\rho_i(\omega_1-\omega_{01})(\omega_h-\omega_{0h,i})}{2\sigma_1\sigma_{h,i}}\right)}\right],\end{align} where the simplified representation has been renormalized.  Note that the single-photon marginal spectra in this model are symmetric about their central frequencies, with $1/\sqrt{e}$ widths of $\sigma$.

The signal photon is then chirped and delayed, which can be represented respectively as a quadratic and linear phase in frequency as \begin{equation}F_i(\omega_1,\omega_h)\mapsto F_i(\omega_1,\omega_h)e^{-i\omega_1\tau+iA_1(\omega_1-\omega_{01})^2},\end{equation} where $\tau$ represents a relative time delay.  The $1/\sqrt{e}$ temporal width of the chirped signal is \begin{equation}\Delta t =\frac{\sqrt{1+16A_1(1-\rho_i^2)\sigma_1^4}}{2\sqrt{1-\rho_i^2}\sigma_1}\label{eq:chirpedwidth}\end{equation} The escort pulse can be represented as a strong coherent state with a field as in \eqref{eq:escortpulse}.  The spectrum of the joint state after upconversion can be found as the convolution of the escort and the input signal spectra~\cite{donohue2015theory} \begin{align}|F_f(\omega_3,\omega_h)|&=\left|\int_{-\infty}^{\infty}\dee\omega_1\,G(\omega_3-\omega_1)\Phi_{SFG}(\omega_1,\omega_3-\omega_1,\omega_3)F_i(\omega_1,\omega_h)\right|\\&= \frac{1}{\sqrt{2\pi\sigma_3\sigma_{h,f}}(1-\rho_i^2)^{1/4}} \exp\left[{\frac{1}{1-\rho_f^2}\left(-\frac{(\omega_3-\omega_{03})^2}{4\sigma_3^2}-\frac{(\omega_h-\omega_{0h,f})^2}{4\sigma_{h,f}^2}-\frac{\rho_f(\omega_3-\omega_{03})(\omega_h-\omega_{0h,f})}{2\sigma_3\sigma_{h,f}}\right)}\right],\end{align} where we are able to re-express the state as a Gaussian since the convolution of two Gaussians is once again a Gaussian.

Even with Gaussian approximations to all the input functions, the theoretical final joint spectral intensity is difficult to express concisely.  For the sake of intuition, we will look at its behaviour under various simplifications.  First, we assume that phasematching is infinitely broad ($\Phi_{SFG}(\omega_1,\omega_3-\omega_1,\omega_3)\approx1$).  As we will see in Supplementary Section~\ref{sec:tunability}, this assumption is not valid, but the phasematching is not so strong that it destroys the intuition gained from the simpler case.  With this simplification, the spectral bandwidth $\sigma_3$ of the output is \begin{equation}\sigma_3=\sqrt{\frac{(2-\rho_i^2)\sigma_1^2\sigma_e^2+\sigma_e^4+(1-\rho_i^2)\sigma_1^4(1+16(A_1+A_e)^2\sigma_e^4)}{\sigma_e^2+(1-\rho_i^2)\sigma_1^2(1+16A_1^2\sigma_1^2\sigma_e^2+16A_e^2\sigma_e^4)}}\label{eq:finalwidth}\end{equation} and the statistical correlation is \begin{equation}\begin{aligned}\rho_f=&-\rho_i\frac{\sqrt{A_1^2(1-\rho_i^2)\sigma_1^2+A_e^2\sigma_e^2}\times \left\{1+16(A_1+A_e)(1-\rho_i^2)\sigma_1^2\left[(A_1+3A_e)(1-\rho_i^2)\sigma_1^2+A_e\sigma_e^2\left(1+16(A_1+A_e)^2(1-\rho_i^2)\sigma_1^4\right)\right]\right\}} {\sqrt{2}\sqrt{\sigma_e^2+2(1-\rho_i^2)\sigma_1^2\left[1+8(A_1+A_e)^2(1-\rho_i^2)\sigma_1^2\sigma_e^2\right]}} \\&\times\frac{\sigma_e^2}{\left\{\sigma_e^2+(1-\rho_i^2)\sigma_1^2\left[1+16A_1^2(1-\rho_i^2)\sigma_1^2\sigma_e^2+16A_e^2\sigma_e^4\right]\right\} \left\{(2-\rho_i^2)\sigma_1^2\sigma_e^2+\sigma_e^4+(1-\rho_i^2)\sigma_1^4\left[1+16(A_1+A_e)^2\sigma_e^4\right]\right\}}\label{eq:finalcorr}\end{aligned}\end{equation}

These expressions are difficult to parse without additional assumptions. Firstly, we look at the limit where the escort has infinite spectral support, $\sigma_e\gg\sigma_1$. As any chirp on a pulse with infinite spectral support stretches it infinitely, this is equivalent to the assumption that the chirped escort is much broader in time than the chirped signal if $|A_e|\neq0$.  In this limit, we find that the width of the upconverted signal is \begin{equation}\lim_{\sigma_e\rightarrow\infty}\sigma_3=\frac{\sqrt{\frac{1}{1-\rho_i^2}+16(A_1+A_e)^2\sigma_1^4}}{4A_e\sigma_1}\overset{LCL}{=}\frac{|A_1+A_e|\sigma_1}{A_e}=M_{spectral}\sigma_1,\end{equation} where $M_{\mathrm{spectral}}=1/M_{\mathrm{temporal}}$ is as defined in Eq.~(2) of the main text, and the statistical correlation is \begin{equation}\lim_{\sigma_e\rightarrow\infty}\rho_f=-\rho_i\frac{4(A_1+A_e)\sigma_1^2\sqrt{1+16(A_1+A_e)^2(1-\rho_i^2)^2\sigma_1^4}}{1+16(A_1+A_e)^2(1-\rho_i^2)\sigma_1^4} \overset{LCL}{=}-\rho_i\frac{A_1+A_e}{|A_1+A_e|} \end{equation}  The simplification on the right-hand-side is once again the large-chirp limit (LCL), where we assume that $16(A_1+A_e)^2(1-\rho_i^2)^2\sigma_1^4\gg1$.  We have also assumed that $A_1\neq -A_e$ in this simplification; if $A_1=-A_e$, the process acts as a time-to-frequency converter, as described in Refs.~\cite{lavoie13comp,donohue2015theory}.  This simplification also relies on the initial state not being perfectly entangled, $|\rho_i|<1$, as the marginal temporal length approaches infinity as the entanglement strengthens (consistent with being generated from truly-continuous-wave pumping). In this large chirp limit, the output waveform is recompressible through spectral phases without need for an additional time lens, and the joint spectrum maintains its degree of correlation.  Were it not for the large chirp limit, some information about the spectrum of the herald would be present in the temporal profile of the upconverted signal, and the statistical correlation of the joint spectra would be reduced. This effect could be corrected with a second time lens forming a complete temporal telescope~\cite{foster2009ultrafast}.

In our experiment, the escort pulse has a bandwidth on the same order as that of the signal, $\sigma_e\sim\sigma_1$.  We next focus on the case studied in our experiment, where $A_e=-A_1/2$.  In the large-chirp limit of this scenario, the output width is \begin{equation}\sigma_3^{(M=-1)}\overset{LCL}{=}\frac{\sigma_1}{\sqrt{\frac{4\sigma_1^2}{\sigma_e^2}+1}},\end{equation} and the statistical correlation is \begin{equation}\rho_f^{(M=-1)}\overset{LCL}{=}\frac{-\rho_i}{\sqrt{\frac{4(1-\rho_i^2)\sigma_1^2}{\sigma_e^2}+1}}.\end{equation}  For both the bandwidth and correlation, the final value has the same absolute value as the input if the spectral support of the escort is sufficient, $\sigma_e\gg4\sigma_1(1-\rho_i^2)$.  Additionally, the sign of the statistical correlation is reversed, $\rho_f=-\rho_i$, consistent with a magnification of -1.  The deviations from this are due to lack of spectral support from the escort, and are exaggerated since the input signal is more strongly chirped than the escort; if the escort is not wider in frequency, it will be shorter in time when chirped and not encompass the full waveform~\cite{donohue2015theory}.  In this case, the escort will act partially as a filter which will degrade the strength of entanglement.

\section{Additional experimental details}\label{sec:TLExpDetails}

\begin{figure}[h!]
  \begin{center}
           \includegraphics[width=1\columnwidth]{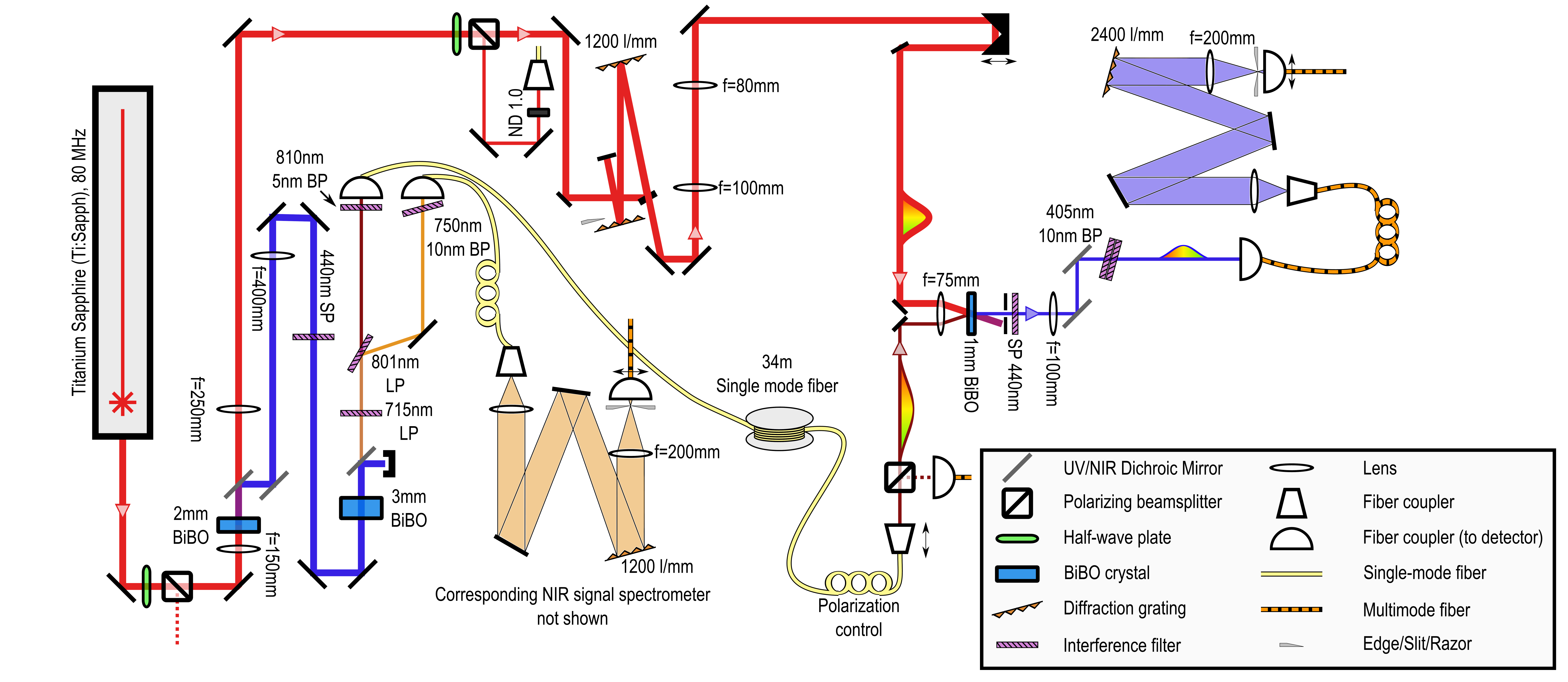}
  \end{center}
 \caption{\textbf{Full-detail experimental setup.}  The experimental setup, as sketched in the main text, is more accurately represented as seen above.  Filters can be longpass (LP), shortpass (SP), or bandpass (BP). Fiber loops with paddles are used to control polarization.}\label{fig:detsetup}
\end{figure}

The experiment uses a titanium sapphire (Ti:Sapph) laser (Coherent Chameleon Ultra II, 80~MHz repetition rate).  The fundamental of the Ti:Sapph was frequency-doubled in 2~mm of type-I phasematched bismuth borate (BiBO) to a second harmonic at $387$~nm with a bandwidth of $0.75$~nm full-width at half-maximum (FWHM). and a power of 935~mW.  The photon pairs were then produced by focusing the second harmonic into 3~mm of type-I BiBO and split with a dichroic mirror, with central wavelengths of 811.0~nm and 740.2~nm for the signal and herald, respectively.

The remaining Ti:Sapph fundamental is re-collimated and used as the escort pulse, which has a central wavelength of $774.6$~nm with a bandwidth of $5.5$~nm FWHM and a power of 945~mW.  Part of the beam was split using a half-wave plate and PBS and delay-matched with the signal photon for alignment.  The rest of the pulse passed through a double-pass grating-based pulse compressor which provides an anti chirp of $A_e=-344\times10^3~\mathrm{fs}^2$~\cite{lavoie13comp}, with a razor blade inserted where the beam was at its widest to effectively act as an adjustable shortpass filter.  The signal photon was chirped in 34~m of single-mode fiber, which provides a chirp of $A_i=696\times10^3~\mathrm{fs}^2$ through material dispersion.  Owing to this fiber delay, the photon originates 13 pulses ahead of the escort.  The chirped signal photon and anti-chirped escort pulse co-propagated with a spatial separation of approximately 9~mm and were then focused into 1~mm of type-I BiBO for sum-frequency generation.  The upconverted beam, with a central wavelength of 396.1~nm, was then re-collimated and the second harmonic of the escort was removed with a pair of bandpass filters.  The upconverted photons were then collected in multimode fiber after passing through a set of polarization optics identical to the herald photon.  The herald photons are detected with Perkin-Elmer SPCM-AQ4C photon counting modules, with detection efficiencies of approximately 50\% near 800~nm.  The upconverted photons were detected with Hamamatsu H10682-210 photon counters, with a quantum efficiency of approximately 30\% near 400~nm.  Coincidence counts were obtained with a window of 3~ns, which is larger than the timing jitter of the electronics and much smaller than the 12.5~ns pulse separation.

The joint spectra were measured using three scanning spectrometers, one for each of the near-infrared (NIR) SPDC photons and one for the upconverted photon.  The beams were expanded to approximately 3.5~mm waist radius and directed to a grating for spectral separation (1200 lines/mm for NIR, 2400 lines/mm for ultraviolet).  The beams were then focused onto a slit and multimode fiber coupler, which move together to measure the full spectrum.  The resolution of spectrometers, measured using emission spectra of a calibration lamp, were found to be $(0.136\pm0.013)$~nm, $(0.148\pm0.012)$~nm, and $(0.0741\pm0.0011)$~nm, for the NIR signal, NIR herald, and upconverted spectrometers, respectively.

The photons were produced through SPDC at a rate of approximately 415,000 coincidence counts per second, with 2.5$\times10^6$ (3.2$\times10^6$)  single-detection events per second for the signal (herald).  After upconversion (but before the spectrometer), approximately 980 coincidence counts (7820 upconverted singles) per second were measured, with approximately 110 (2820) of those being background, of which the most significant source was the second harmonic of the escort pulse.  This second harmonic background is spectrally resolvable from the upconverted photons, and thus does not have a significant effect on the joint spectral measurements.

See Fig.~S-\ref{fig:detsetup} for a detailed map of the experimental setup. See Table~S-\ref{tab:jointspecFull} for a collection of fit parameters of the joint spectra seen in Fig.~3 of the main text.

\begin{table}[h!]\centering
\begin{tabular}{|M{0.255\columnwidth}|M{0.175\columnwidth}|M{0.175\columnwidth}|M{0.175\columnwidth}|M{0.175\columnwidth}|}
\hline
 \multirow{2}{*}{Property} & \multicolumn{2}{c|}{Input} & \multicolumn{2}{c|}{Output} \\ \cline{2-5}
 & Raw & Deconvolved & Raw & Deconvolved \\
 \hline\hline
 Signal central wavelength & \multicolumn{2}{c|}{$(811.006\pm0.003)$~nm}& \multicolumn{2}{c|}{$(396.113\pm0.004)$~nm}\\ \hline
 Signal bandwidth & $(4.047\pm0.005)$~nm & $(4.034\pm0.006)$~nm & $(0.621\pm0.010)$~nm&  $(0.60\pm0.01)$~nm\\ \hline\hline
 Herald central wavelength & \multicolumn{2}{c|}{$(740.194\pm0.003)$~nm} & \multicolumn{2}{c|}{$(740.126\pm0.018)$~nm} \\ \hline
 Herald bandwidth & $(3.733\pm0.005)$~nm & $(3.716\pm0.006)$~nm & $(2.50\pm0.04)$~nm & $(2.47\pm0.04)$~nm\\ \hline\hline
 Correlation $\rho$ & $-0.97024\pm0.00015$  & $-0.9776\pm0.0009$ & $0.863\pm0.003$ & $0.909\pm0.005$\\ \hline
 Schmidt rank $K$ & $4.13\pm0.01$ & $4.75\pm0.1$ & $1.98\pm0.03$ & $2.39\pm0.06$\\ \hline
 Joint energy uncertainty & $(0.334\pm0.001)$~THz & $(0.333\pm0.001)$~THz & $(0.468\pm0.008)$~THz & $(0.455\pm0.008)$~THz\\ \hline
\end{tabular}
\caption{\textbf{Complete fit joint spectral parameters.}  Selected properties of the Gaussian fits to the joint spectra seen in Fig.~3 of the main text are given above.  The deconvolved values are corrected for the finite resolution of the spectrometers used.  All widths are reported full-width at half-maximum.  We also calculate the Schmidt rank of the state under the assumption that the state is both pure and that any frequency-dependent phases do not affect the Schmidt decomposition, which is related to the statistical correlation as ${K=(1-\rho^2)^{-1/2}}$ from the purity of the partial trace.  The joint energy uncertainty is defined as the width of the semi-minor axis of an elliptical Gaussian fit. The error bars are determined from Monte Carlo simulation of the data with the assumption of Poissonian count statistics. \label{tab:jointspecFull} }
\end{table}

\section{Tunability of the joint spectrum}\label{sec:tunability}

\begin{table}[b!]\centering
\begin{tabular}{|M{0.15\columnwidth}|M{0.15\columnwidth}||M{0.16\columnwidth}|M{0.16\columnwidth}||M{0.16\columnwidth}|M{0.16\columnwidth}|}
\hline
 \multicolumn{2}{|c||}{} & \multicolumn{2}{c||}{Upconverted signal tunability} & \multicolumn{2}{c|}{Herald tunability} \\
 \hline
 Effective crystal length $L$ & Effective escort bandwidth $\sigma_e$ & $\delta\omega_{03}/2\pi$ & $\delta\lambda_{03}$ & $\delta\omega_{0h}/2\pi$ & $\delta\lambda_{0h}$\\
 \hline\hline
 \multicolumn{2}{|c||}{Measured} & 0.14~THz/ps & 0.071~nm/ps & -0.097~THz/ps & -0.18~nm/ps\\
 \hline\hline
 $L_{meas.}$ & $\sigma_{e,meas.}$ & 0.12~THz/ps & 0.061~nm/ps & -0.12~THz/ps & -0.22~nm/ps\\
  \hline
 $0.65\times L_{meas.}$ & $\sigma_{e,meas.}$ & 0.14~THz/ps & 0.071~nm/ps & -0.099~THz/ps & -0.18~nm/ps\\
   \hline
 $0$ & $\infty$ & 0.23~THz/ps & 0.121~nm/ps & 0~THz/ps & 0~nm/ps\\
 \hline
 $0$ & $\sigma_{e,meas.}$ & 0.16~THz/ps & 0.081~nm/ps & -0.080~THz/ps & -0.14~nm/ps\\
 \hline
  $L_{meas.}$ & $\infty$ & 0.12~THz/ps & 0.063~nm/ps & -0.11~THz/ps & -0.21~nm/ps\\
 \hline

\end{tabular}
\caption{\textbf{Expected tunabilities with various assumptions.}  The expected tunabilities of the central frequency for the upconverted signal and herald are given for a variety of assumptions, and can be compared with the tunabilities measured in the experiment, shown in Fig.~S-\ref{fig:tunable}f.  The experimental results are in agreement with the theory when the length is scaled to 65\% of its physical value, likely due to the non-collinear geometry of the interaction.  The ideal-world theory corresponds to where the phasematching is nonrestrictive ($L=0$) and the escort has infinite spectral support ($\sigma_e\rightarrow\infty$).  The experimental values are $L_{meas.}=1$~mm and $\sigma_{e,meas.}=7.38\times10^{12}\,\mathrm{s}^{-1}=5.53$~nm~FWHM.\label{tab:tunability} }
\end{table}

\begin{figure}[h!]
  \begin{center}
           \includegraphics[width=1\columnwidth]{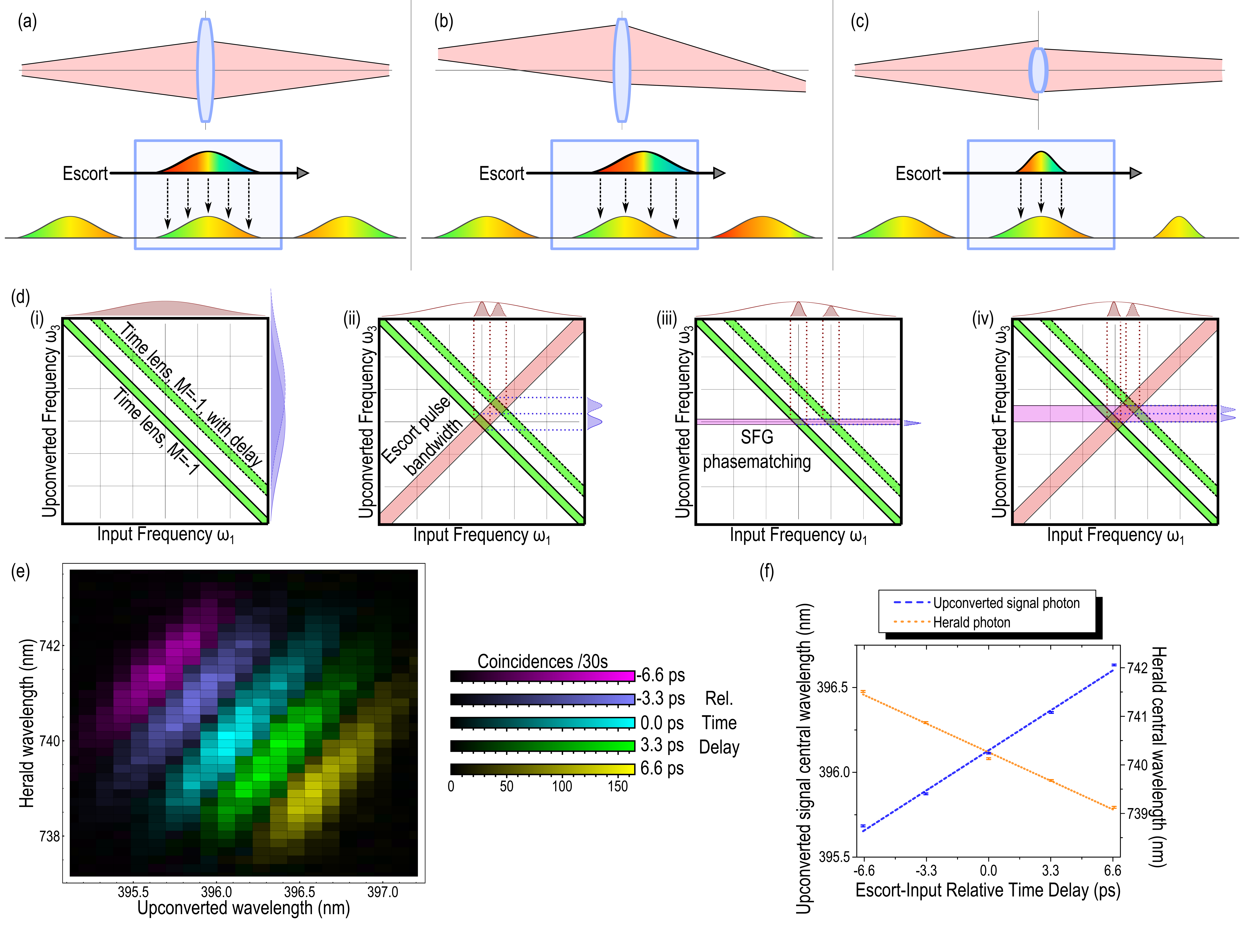}
  \end{center}
 \caption{\textbf{Joint spectrum tunability.} (a) The effect of the escort pulse on the signal in an upconversion time lens can be understood in analogy with the action of a lens in a spatial imaging system. (b)  In an analogous fashion to how shifting the spatial center of a lens causes the deflection of a beam in space, a relative time delay on the escort pulse of an upconversion time lens causes a shift in the center frequency of the upconverted light.  (c)  If the spectral support of the escort pulse is limited, the upconverted light will have a narrower spectrum and longer temporal duration than the ideal scenario, analogous to a lens with limited clear aperture.  (d)  The reason both the output signal and herald central wavelengths may change can be understood by taking into account the effect of the chirp, finite escort bandwidth, and finite phasematching.  The four plots shown give the mapping of an input frequency to an output frequency in the time-lens scenario; in the case where the input signal is frequency anti-correlated with a herald, the expected value of herald in the joint state will shift depending on what portions of the input were successfully upconverted. (i) The chirps in the pulse, assuming a magnification of $M=-1$, flip an input frequency blue-shifted relative to its centre to one red-shifted of its new centre.  Introducing a time delay effectively changes the centre, as can be visualized with two offset negatively sloped diagonal lines. In the absence of other effects, the input spectra is accepted in full, and thus the herald spectra will be unchanged when measured in coincidence. (ii) The sum-frequency process must conserve energy, and limiting the bandwidth of the escort pulse will enforce stricter conservation.  If we consider the limit where the pulses are stretched well beyond their Fourier-limited widths by the chirp, this will result in the escort pulse acting as a filter, causing corresponding shifts in the signal and herald central frequencies when measured in coincidence.  (iii)  In the limit when the escort and input signal are degenerate in the sum-frequency process, the phasematching is roughly a restriction on the upconverted frequencies accessible.  As changing the time delay changes which input frequency converts to a specific upconverted frequency, the herald spectrum may shift while the upconverted spectrum remains stable in this limit.  (iv)  As we see both central frequencies change as a function of delay but with difference slopes in terms of energy, we conclude that we observe a mixture of these competing effects. (e) Five measured joint spectra are shown with varied relative time delays between the escort pulse and input signal, reprinted from Fig.~4 of the main text.  The shift in both the herald and output signal central wavelength is clearly apparent, but positive correlations are maintained throughout.  (f) The fit upconverted (herald) central wavelength is seen to increase (decrease) as the relative delay is increased, with a best-fit slope of $0.071$~nm/ps ($-0.179$~nm/ps).}\label{fig:tunable}
\end{figure}

When a spatial lens is off-center relative to the incoming beam, the focus of the beam is translated, as in Fig.~S-\ref{fig:tunable}a-b.  Analogously, when a relative delay exists between the escort pulse and the input signal in an upconversion time lens, the central frequency of the upconverted signal will shift; a similar behaviour occurs in chirped-pulse bandwidth compression~\cite{lavoie13comp}.  When we neglect the effects of limited escort bandwidth and restrictive phasematching (i.e. $\sigma_e\rightarrow\infty$ and $\Phi_{SFG}(\omega_1,\omega_3-\omega_1,\omega_3)\approx1$) and take the large-chirp limit, the central frequency of the upconverted signal from an $M=-1$ time lens can be found to be simply \begin{equation}\omega_{03}^{(ideal)}=\omega_{01}+\omega_{0e}+\frac{\tau}{A_1},\end{equation} where $\tau$ is a relative delay applied to the signal, $\omega_{01}$ and $\omega_{0e}$ are the central frequencies of the input signal and escort pulse respectively, and $A_1$ is the chirp on the input signal (the escort chirp $A_e$ is assumed to be exactly $-\frac{1}{2}A_1$, as needed for $M=-1$).  The shift in herald center frequency, $\delta\omega_{0h,f}$, is zero in this limit.

However, once limited escort bandwidth and phasematching are taken into consideration, this approximation breaks down. The effect of each of these imperfections is described graphically in Fig.~S-\ref{fig:tunable}d.  In the large-chirp limit, we can look at the limit where $\sigma_e\ll\sigma_1$, i.e. the escort is much narrower spectrally than the input signal.  In this case, the chirped escort is much narrower in time than the chirped photon, and the upconversion acts as a filter, as shown in Fig.~S-\ref{fig:tunable}c.  The central frequency shifts are then \begin{align}\lim_{\sigma_e\ll\sigma_1}\delta\omega_{03}&\overset{LCL}{=}\frac{\tau}{2A_1}\\ \lim_{\sigma_e\ll\sigma_1}\delta\omega_{0h,f}&\overset{LCL}{=}\frac{\rho\sigma_{h,i}\tau}{2A_1\sigma_1}.\end{align} If $\sigma_{h,i}\approx\sigma_1$ and the initial photon pair has strong frequency anti-correlations $\rho_i\approx-1$, the shift in each is simply opposite, which is due to selecting different parts of the input rather than upconverting the entire pulse.

We next look at the limit where the phasematching is restrictive.  As we experimentally consider a sum-frequency process which is nearly degenerate between the escort and input signal, we make the approximation that the group velocities in the sum-frequency media of the two inputs are roughly equal.  In this case, the phasematching function $\Phi_{SFG}(\omega_1,\omega_e,\omega_3)\approx\Phi_{SFG}(\omega_3)$, which we approximate to a Gaussian which grows tighter as the effective crystal length $L$ is increased.  In the large-chirp long-crystal limit, the upconverted signal is effectively untunable as phasematching restricts it to a specific output waveform, but the herald central frequency is found to be \begin{align} \lim_{LCL}\lim_{L\rightarrow\infty}\delta\omega_{0h,f}&=\frac{\rho\sigma_{h,i}\tau}{A_1\sigma_1}.\end{align}  The herald changes since the phasematching only allows one frequency to be upconverted to, and the chirps applied force every interaction to be nearly monochromatic.  Delaying the input signal changes which combination of frequencies from the escort and signal will combine to the accepted upconversion frequency, and will change which part of the input signal is upconverted.


We took five joint spectra as the delay between the escort and input signal was altered, as seen in Fig.~S-\ref{fig:tunable}e, and saw a shift in both the herald and signal central frequencies, as seen in Fig.~S-\ref{fig:tunable}f.  Given our chirp parameter of $A_1=696\times10^3\mathrm{fs}^2$, we would expect a tunability of $0.23\frac{THz}{ps}=0.12\frac{nm}{ps}$ in an idealized case, but instead measure $0.14\frac{THz}{ps}=0.071\frac{nm}{ps}$.  The herald central frequency is also shifted, which would not occur ideally, with a slope of $-0.097\frac{THz}{ps}=0.18\frac{nm}{ps}$.  Our results our inconsistent with the idealized theory, as mentioned above, but are relatively consistent with a process modelled by the bandwidths we measured and a shortened effective crystal length, as seen in Table S-\ref{tab:tunability}.  The shortened effective crystal length is expected as the signal photon and escort are combined non-collinearly in the upconversion medium and therefore do not interact for the entire length of the crystal.

\end{document}